\documentstyle[epsf,prd,aps,floats]{revtex}

\input{psfig.tex}

\begin{document}
\twocolumn[\hsize\textwidth\columnwidth\hsize\csname@twocolumnfalse\endcsname

\author{John D. Barrow$^1$, H\aa vard Bunes Sandvik$^2$, and Jo\~{a}o Magueijo$^2$}
\address{$^1$DAMTP, Centre for Mathematical Sciences,\\
Cambridge University, Wilberforce Rd., \\
Cambridge CB3 0WA, UK\\
$^2$Blackett Laboratory, Imperial College, \\
Prince Consort Rd., London, SW7 2BZ, UK}
\title{Anthropic Reasons for Non-zero Flatness and Lambda}
\date{}
\maketitle

\begin{abstract}
In some cosmological theories with varying constants there are
anthropic reasons why the expansion of the universe must not be
too {\it close} to flatness or the cosmological constant too close
to zero. Using exact theories which incorporate time-variations in
$\alpha $ and in $G$ we show how the presence of negative spatial
curvature and a positive cosmological constant play an essential
role in bringing to an end variations in the scalar fields driving
time change in these 'constants' during any dust-dominated era of
a universe's expansion. In spatially flat universes with $\Lambda
=0$ the fine structure constant grows to a value which makes the
existence of atoms impossible.
\end{abstract}

\pacs{PACS Numbers: *** }
]

\section{Introduction}

The collection of considerations now known as the Anthropic Principles
emerged from attempts by Whitrow \cite{whit} to understand why it is
unsurprising that we find space to have three dimensions, and by Dicke \cite
{dic} to understand the inevitability of Dirac 'large number' coincidences
in cosmology. Dicke recognised that it was unnecessary to introduce the idea
of a time-varying gravitational constant in order to understand why we could
not fail to observe that the number of protons in the observable universe is
of order the square of the ratio of electromagnetic to gravitational force
strengths. Subsequently, Dicke inspired a detailed observational and
theoretical investigation of gravity theories in which the Newtonian
gravitational constant becomes a space-time variable. he was partly
motivated by apparent discrepancies between the predictions of standard
general relativity and observations of the perihelion precession of Mercury.
These discrepancies were subsequently ascribed to errors in the measurements
of the shape and diameter of the Sun created by solar surface activity \cite
{will}.

There have been many investigations of the apparent coincidences that allow
complexity to exist in the universe (see \cite{cr,btip,teg,hog}). Typically,
they examine the stability of life-supporting conditions to small (or large)
perturbations to the values of constants of Nature or to quantities fixed by
cosmological 'initial' conditions at $t=0$ or $t=-\infty $. These in turn
divide into studies of two sorts: first, those in which the hypothetical
changes introduced to the 'constants' are self-consistently permitted by the
cosmological or physical theory employed; and second, those in which they
are not. An investigation of the first kind might be one in which the
cosmological initial conditions were enlarged to allow anisotropies or the
possibility of a significant deviation from flatness. An investigation of
the second type might note that a change in the observed value of the
electron to proton mass ratio to another fixed value would make it difficult
to produce ordered molecular structures. Studies of universes in which
traditional 'constants' of Nature are changed are restricted by the lack of
self-consistent theories which allow all these possible changes to be
accommodated. Without them, it is impossible to determine the possible
knock-on effects of varying one constant on others.

There are some exceptions. Varying gravitation 'constant', $G$,
(or dimensionless constants formed with it like $Gm^2/hc$ for any
mass $m$), can be studied using scalar-tensor gravity theories
\cite{BD}. A varying fine structure 'constant' can be studied
using the theory of Bekenstein and Sandvik, Barrow and Magueijo
(BSBM) \cite{bek2}, \cite{sbm}. Moreover, the formulation of
physical theories whose true constants inhabit more than three
space dimensions provides a framework for the rigorous study of
the simultaneous variation of their three-dimensional counterparts
\cite{dims}, \cite{marc}, \cite{drink}. Recently there has also
been much interest in theories where a variation in the fine
structure constant is due to a change in the light propagation
speed\cite{moffat93,am,ba}. In another paper we propose various
methods for experimentally distinguishing between these different
theories\cite{mbs}.

Observational evidence for a variation in a traditional constant
can be found without the need for a self-consistent theory of its
variation simply by demonstrating incompatibility with the
predictions of the standard theory. The most observationally
sensitive 'constant' is the electromagnetic fine structure
constant, $\alpha \equiv e^2/\hbar c$, and recent observations
motivate the formulation of varying-$\alpha $ theories. The new
many-multiplet technique of Webb et al, \cite{murphy},
\cite{webb}, exploits the extra sensitivity gained by studying
relativistic transitions to different ground states using
absorption lines in quasar (QSO) spectra at medium redshift. It
maximises the information extracted from the data set and has
provided the first evidence that the fine structure constant might
change with cosmological time\cite{murphy,webb,webb2}. The trend
of these results is that the value of $\alpha $ was lower in the
past, with $\Delta \alpha /\alpha =-0.72\pm 0.18\times 10^{-5}$
for $z\approx 0.5-3.5.$ Other investigations
\cite{avelino,battye,avelino2}have claimed preferred non-zero
values of $\Delta \alpha <0 $ to best fit the cosmic microwave
background (CMB) and Big Bang Nucleosynthesis (BBN) data at
$z\approx 10^3$ and $z\approx 10^{10}$ respectively but appeal to
much larger variations. We have shown that the simplest theory
which joins varying $\alpha $ to general relativity via the
propagation of a scalar field can explain these observations
together with the lack of evidence for a similar level of
variation locally, 2 billion years ago, or at very high redshifts,
$z\geq 10^3$ . In this paper we will show how this theory also
provides some novel anthropic perspectives on the evolution of our
universe or others.

There have been several studies, following Carter, \cite{car} and Tryon \cite
{try}, of the need for life-supporting universes to expand close to the
'flat' Einstein de Sitter trajectory for long periods of time. This ensures
that the universe cannot collapse back to high density before galaxies,
stars, and biochemical elements can form by gravitational instability, or
expand too fast for stars and galaxies to form by gravitational instability
(see also \cite{ch}, \cite{barqj} and \cite{btip}). Likewise, it was pointed
out by Barrow and Tipler, \cite{btip} that there are similar anthropic
restrictions on the magnitude of any cosmological constant, $\Lambda $. If
it is too large in magnitude it will either precipitate premature collapse
back to high density (if $\Lambda <0$) or prevent the gravitational
condensation of any stars and galaxies (if $\Lambda >0$). Thus existing
studies provide anthropic reasons why we can expect to live in an old
universe that is neither too far from flatness nor dominated by a much
stronger cosmological constant than observed ($\left| \Lambda \right| \leq
10\left| \Lambda _{obs}\right| $).

Inflationary universe models provide a possible theoretical explanation for
proximity to flatness but no explanation for the smallness of the
cosmological constant. Varying speed of light theories \cite
{moffat93,am,ba,bm} offer possible explanations for proximity to flatness
and smallness of a classical cosmological constant (but not necessarily for
one induced by vacuum corrections in the early universe). Here, we shall
show that if we enlarge our cosmological theory to accommodate variations in
some traditional constants then{\it \ it appears to be anthropically
disadvantageous for a universe to lie too close to flatness or for the
cosmological constant to be too close to zero}. This conclusion arises
because of the coupling between time-variations in constants like $\alpha $
and the curvature or $\Lambda $, which control the expansion of the
universe. The onset of a period of $\Lambda $ or curvature domination has
the property of dynamically stabilising the constants, thereby creating
favourable conditions for the emergence of structures. This point has been
missed in previous studies because they have never combined the issues of $%
\Lambda $ and flatness and the issue of the values of constants. By coupling
these two types of anthropic considerations we find that too little $\Lambda
$ or curvature can be as poisonous for life as too much.

\section{Time variation of $\alpha $}

First, consider a simple theory with varying $\alpha \equiv
e^{2\psi }/\hbar c$ where $\psi $ is a scalar field that can vary
in space and time. A generalisation of the scalar theory proposed
by Bekenstein \cite{bek2} described in ref. \cite{sbm} to include
the gravitational effects of $\psi $ gives the field equations
\begin{equation}
G_{\mu \nu }=8\pi G\left( T_{\mu \nu }^{matter}+T_{\mu \nu }^\psi +T_{\mu
\nu }^{em}e^{-2\psi }\right) ,
\end{equation}
and the $\psi $ field obeys the equation of motion
\begin{equation}
\Box \psi =\frac 2\omega e^{-2\psi }{}{\cal L}_{em}.
\label{boxpsi}
\end{equation}
We have defined the coupling constant $\omega = (\hbar c)/ l^2$,
where $l$ is the length scale down to which the theory is
accurately coloumbic. It is clear that ${\cal L}_{em}$ vanishes
for a sea of pure radiation since then ${\cal L}_{em}=(E^2 -
B^2)/2=0$. We therefore expect the variation in $\alpha$ to be
driven by electrostatic and magnetostatic energy-components rather
than electromagnetic radiation. In order to make quantitative
predictions we need to know how much of the non-relativistic
matter contributes to the RHS of Eqn.~(\ref{boxpsi}). This is
parametrised by $\zeta \equiv {\cal L} _{em}/\rho$, where $\rho$
is the energy density , and for baryonic matter ${\cal
L}_{em}=E^2/2$. In previous papers \cite{sbm,bsm} we showed how
the cosmological value of $\zeta$ (denoted $\zeta_m$) is largely
determined by the nature of dark matter.  To accommodate for a
lower $\alpha$ in the past, as preferred by the data, the dark
matter constituents need to have high magnetostatic energy content
(One possible contender would be superconducting cosmic strings
which have $\zeta_m \sim -1$). In line with our recent work and
the observational data we will in this paper confine ourselves to
negative values of $\zeta_m$.

Assuming a homogeneous and isotropic Friedmann metric with expansion scale
factor $a(t)$ and curvature parameter $k$ we obtain the field equations ($%
c\equiv 1$)
\begin{equation}
\left( \frac{\dot a}a\right) ^2=\frac{8\pi G}3\left( \rho _m\left(
1+|\zeta_m| e^{-2\psi }\right) +\rho _re^{-2\psi }+\frac \omega
2\dot \psi ^2+\rho _\Lambda \right) -\frac k{a^2},  \label{fried1}
\end{equation}
where the cosmological vacuum energy $\rho _\Lambda $ is a constant that is
proportional to the cosmological constant $\Lambda \equiv 8\pi G\rho
_\Lambda $. For the scalar field we have
\begin{equation}
\ddot \psi +3H\dot \psi =-\frac 2\omega e^{-2\psi }\zeta_m \rho _m
\label{psiddot}
\end{equation}
where $H\equiv \dot a/a.$ The conservation equations give for the
non-interacting radiation, and matter densities $\rho _r$ $\propto e^{2\psi
}a^{-4}$and $\rho _m\propto a^{-3},$ respectively. This theory enables the
cosmological consequences of varying $\alpha $, to be analysed
self-consistently rather than by changing the constant value of $\alpha $ in
the standard theory, as in the original proposals made in response to the
large numbers coincidences \cite{gam}.

\begin{figure}[ht]
\psfig{file=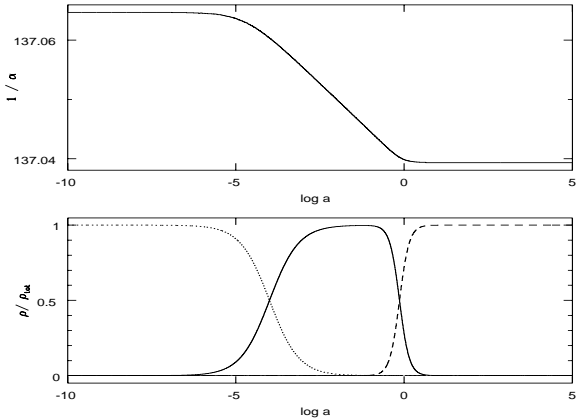,height=6cm,width=8cm} \caption{The top
plot shows the change in alpha throughout the dust epoch ends as
lambda takes over the expansion. The lower plot shows the
radiation (dotted), dust (solid) and lambda (dashed) densities as
fractions of the total energy density.} \label{lambdafig}
\end{figure}

The cosmological behaviour of the solutions to these equations was studied
by us \cite{sbm}, \cite{bsm} for the $k=0$ case and is shown in Figure (\ref
{lambdafig}). The evolution of $\alpha $ is summarised as follows:

1. During the radiation era $\alpha $ is constant
and $ a(t)\sim t^{1/2}$. It
increases in the dust era, where $a(t)\sim t^{2/3}$, until the
cosmological constant starts to accelerate the universe, $a(t)\sim
\exp [\Lambda t/3],$ after which $\alpha $ asymptotes rapidly to a
constant, see fig.(\ref{lambdafig})

\begin{figure}[ht]
\psfig{file=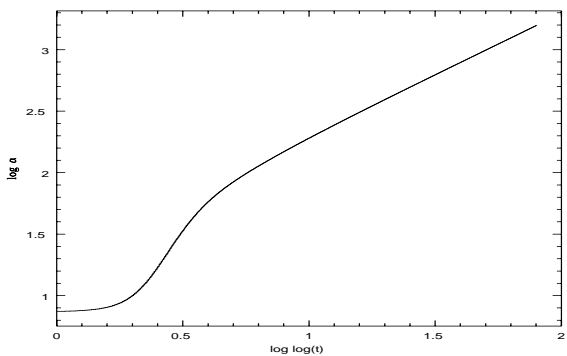,height=5cm,width=8cm} \caption{$\psi \propto
\ln \alpha$ changes as $\log (2N \log t)$ in the dust era.}
\label{dustfig}
\end{figure}

2. If we set the cosmological constant equal to zero then, during
the dust era, $\alpha $ will increase indefinitely. The increase
however, is very slow with a late-time solution for $\psi $
proportional to $\log (2N \log (t))$, see fig.(\ref{dustfig}).  N
is defined as $N \equiv -2 \zeta_m / \rho_m a^3$, a positive
constant since we have confined ourselves to $\zeta_m < 0$.

\begin{figure}[tbp]
\psfig{file=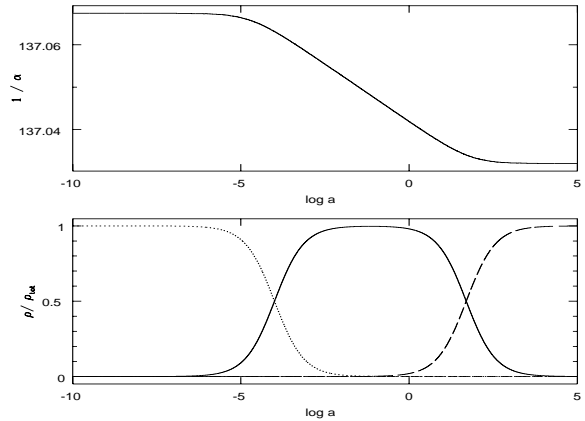,height=6cm,width=8cm} \caption{Top: The
change in alpha comes to an end as curvature takes over the
expansion. The bottom graph again shows the different constituents
of the universe as a function of the scale factor.}
\label{curvfig}
\end{figure}

3. If we set the cosmological constant equal to zero and introduce a
negative spatial curvature ($k<0$) then $\alpha $ increases only during the
dust-dominated phase, where $a(t)\sim t^{2/3}$, but tends to a constant
after the expansion becomes curvature dominated, with $a(t)\sim t$. This
case is illustrated in fig.(\ref{curvfig}).

From these results it is evident that non-zero curvature or cosmological
constant brings to an end the increase in the value of $\alpha $ that occurs
during the dust-dominated era\footnote{%
In some Friedmann universes with initial conditions unlike our own there can
be power-law growth of $\alpha $ during the radiation era,  \cite{bsm}. In
such universes the same general effects of negative curvature and positive $%
\Lambda $ are seen. They still halt any growth in $\alpha (t).$
Our initial conditions are chosen so as to give a present day
value of $\alpha \approx 1/137$. The initial value of alpha would
have to be several orders of magnitude lower in order to obtain
the power-law growth}. Hence, if the spatial curvature and
$\Lambda $ are too small it is possible for the fine structure
constant to grow too large for biologically important atoms and
nuclei to exist in the universe. There will be a time in the future when $%
\alpha $ reaches too large a value for life to emerge or persist. The closer
a universe is to flatness or the closer $\Lambda $ is to zero so the longer
the monotonic increase in $\alpha $ will continue, and the more likely it
becomes that life will be extinguished. Conversely, a non-zero positive $%
\Lambda $ or a non-zero negative curvature will stop the increase of $\alpha
$ earlier and allow life to persist for longer. If life can survive into the
curvature or $\Lambda $-dominated phases of the universe's history then it
will not be threatened by the steady cosmological increase in $\alpha $
unless the universe collapses back to high density.

\section{Anthropic Limits on $\alpha $}

We have seen that varying-$\alpha $ cosmologies with zero curvature and $%
\Lambda $ lead to a monotonic increase in $\alpha $ with time. Here we
summarise the principal upper limits on $\alpha $ that are needed for atomic
complexity and stars to exist. There are a variety of constraints on the
maximum value of the fine structure compatible with the existence of
nucleons, nuclei, atoms and stars under the assumption that the forms of the
laws of Nature remain the same. The running of the fine structure constant
with energy due to vacuum polarisation effects leads to an exponential
sensitivity of the proton lifetime with respect to the low-energy value of $%
\alpha $ with $t_{pr}\sim \alpha ^{-2}\exp (\alpha ^{-1})m_{pr}^{-1}\sim
10^{32}yrs.$ In order that the lifetime be less than the main sequence
lifetime of stars we have $t_{pr}<(Gm_{pr}^2)^{-1}m_{pr}^{-1}$ which implies
that $\alpha $ is bounded above by $\alpha <1/80$ approximately \cite{ell}.

The stability of nuclei is controlled by the balance between nuclear binding
and electromagnetic surface forces \cite{dav}. A nucleus ($Z,A$) will be
stable if $Z^2/A<49(\alpha _s/0.1)^2(1/137\alpha ).$ In order for carbon $%
(Z=6)$ to be stable we require $\alpha <16(\alpha _s/0.1)^2.$ Detailed
investigations of the nucleosynthesis processes in stars have shown that a
change in the value of $\alpha $ by $4\%$ shifts the key resonance level
energies in the carbon and oxygen nuclei which are needed for the production
of a mixture of carbon and oxygen from beryllium plus helium-4 and carbon-12
plus helium-4 reactions in stars \cite{hoy,csoto}. These upper bounds on $%
\alpha $ are model independent and were considered in more detail in refs.
\cite{btip}, \cite{cr} and \cite{teg}. However, sharper limits can be found
by using our knowledge of the stability of matter derived from analysis of
the Schr\"odinger equation. Stability of matter with Coulomb forces has been
proved for non-relativistic dynamics, including arbitrarily large magnetic
fields, and for relativistic dynamics without magnetic fields. In both cases
stability requires that the fine structure constant be not too large.

The value of $\alpha $ controls atomic stability\footnote{%
Note that if the electron mass and velocity of light are varied along with
the value of $\alpha $ then the eigenvalues of the non-relativistic
Schr\"odinger equation can remain invariant and atomic structure is
unchanged \cite{btip}. Here, we break the scale invariance by varying only $%
\alpha .$}. If $\alpha $ increases in value then the innermost Bohr orbital
contracts and electrons will eventually fall into the nucleus when $\alpha
>Z^{-1}m_{pr}/m_e.$ As $\alpha $ increases, atoms all become relativistic
and unstable to pair production. In order that the electromagnetic repulsion
between protons does not exceed nuclear strong binding $e^2/r_n<\alpha m_\pi
$ is needed and so we require $\alpha <1/20.$ It is also known that atomic
instability of atoms with atomic number $Z$ occurs in the relativistic
Schr\"odinger equation when the fine structure constant is increased in
value to $\alpha =\frac 2{\pi Z}.$ However, when the many-electron and
many-nucleon problem is examined with the relativistic Schr\"odinger theory
there is a bound on $\alpha $ for stability that is independent of $Z$ \cite
{lieb}. If $\alpha <1/94$ then stability occurs all the way up to the
critical value $\alpha =\frac 2{\pi Z}$, whereas if $\alpha $ $>128/15\pi $
the 'atomic' system is unstable for all values of $Z$. In the presence of
arbitrarily large magnetic fields, which aid binding by creating a
two-dimensional form for the potential, matter composed of electrons and
nuclei is known to be unstable if $\alpha $ or $Z$ is too large: matter is
stable if $\alpha $ $<0.06$ and $\alpha $ $<0.026(6/Z)^{1/2}$,\cite{lieb2},%
\cite{seid}.

If stars are to exist, their centres must be hot enough for thermonuclear
reactions to occur. This requires $\alpha $ to be bounded above by $\alpha
^2<20m_e/m_{pr}.$ Carter has also pointed out the existence of a very
sensitive condition $\alpha ^{12}\sim \left( m_e/m_{pr}\right) ^4Gm_{pr}^2\
, $that must be met if stars are to undergo a convective phase, although
this stringent condition no longer seems to be essential for planetary
formation \cite{car}.

The results collected above show that there are a number of general {\it %
upper limits} on the value of $\alpha $ if atoms, molecules, and
biochemistry are to exist. These bounds do not involve the gravitation
constant explicitly. Other astrophysical upper bounds on $\alpha $ exist in
order that stars be able to form but these involve the gravitational
constant.

\section{Time variation of $G$}

{\em \ }A similar trend can be found in relativistic cosmologies in
scalar-tensor gravity theories. Consider the paradigmatic case of
Brans-Dicke (BD) theory to fix ideas. The form of the general solutions to
the Friedmann metric in BD theories are fully understood \cite{bar},\cite
{fink}. The general solutions begin at high density dominated by the BD
scalar field $\phi \sim G^{-1}$ and approximated by the vacuum solution. At
late times they approach particular exact power-law solutions for $a(t)$ and
$\phi (t)$ and the evolution is 'Machian' in the sense that the cosmological
evolution is driven by the matter content rather than by the kinetic energy
of the free $\phi $ field. There are three essential field equations for the
evolution of $\phi $ and $a(t)$ in a BD universe
\begin{figure}[tbp]
\psfig{file=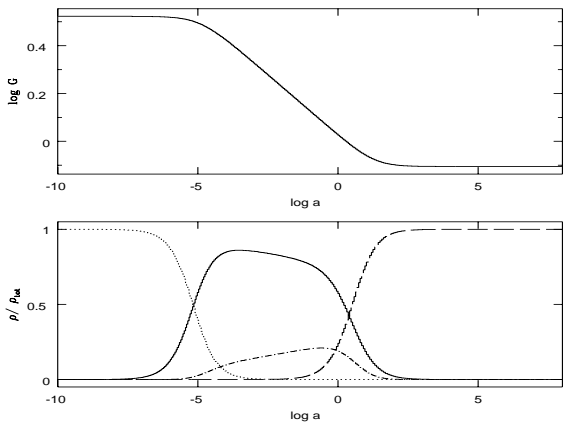,height=6cm,width=8cm}
\caption{Top plot shows cosmological evolution of Brans-Dicke theory, with $%
\omega =10$, from radiation domination into dust domination and through to
curvature driven expansion. Lower plot shows radiation (dotted) , dust
(solid) and curvature (dashed) energies, as well as the scalar field energy
(combined), as a fraction of the total energy density. }
\end{figure}
\begin{figure}[tbp]
\psfig{file=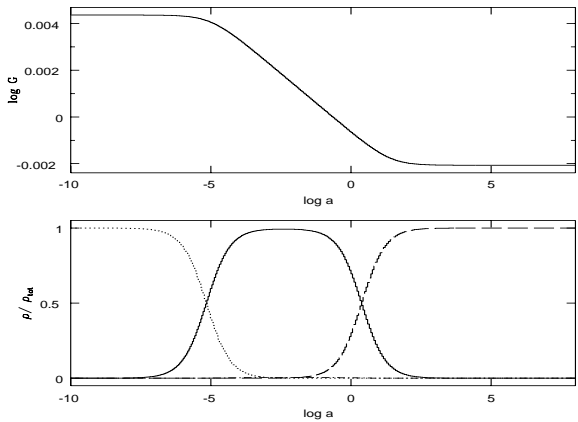,height=6cm,width=8cm} \caption{Similar
evolution of Brans-Dicke theory with $\omega =1000$.}
\end{figure}

\begin{eqnarray*}
3\frac{\dot a^2}{a^2} &=&\frac{8\pi \rho }\phi -3\frac{\dot a^{\ }\dot \phi
}{a\ \phi }+\frac{\omega _{BD}}2\frac{\dot \phi ^2}{\phi ^2}-\frac k{a^2} \\
\ddot \phi +3\frac{\dot a}a\dot \phi &=&\frac{8\pi }{3+2\omega }(\rho -3p) \\
\dot \rho +3\frac{\dot a}a(\rho +p) &=&0
\end{eqnarray*}
Here, $\omega _{BD}$ is the BD constant parameter and the theory reduces to
general relativity in the limit $\omega _{BD}\rightarrow \infty $ and $\phi
=G^{-1}\rightarrow $constant.

In the radiation era the scale factor approaches the standard general
relativistic behaviour for large times:

\begin{equation}
a(t)\sim t^{1/2};\hspace{1.0in}G=constant  \label{r}
\end{equation}
After the dust density dominates the dynamics the expansion approaches a
simple exact solution with

\begin{equation}
a(t)\propto t^{(2-n)/3};\hspace{1.0in}G\propto t^{-n},  \label{dust}
\end{equation}
which continues until the curvature term takes over the expansion. Here, $n$
is related to the constant Brans-Dicke $\omega _{BD}$ parameter by

\begin{equation}
n\equiv \frac 2{4+3\omega _{BD}}  \label{n}
\end{equation}
and the usual general relativistic Einstein de Sitter universe is obtained
as $\ \omega _{BD}\rightarrow \infty $ and $n\rightarrow 0.$ If the universe
is open, ($k=-1$), then the negative curvature will eventually dominate the
gravitational effects of the dust and then the BD model approaches the
general relativistic Milne model with constant $G$

\begin{equation}
a(t)\propto t;\hspace{1.0in}G=constant  \label{mil}
\end{equation}


Again, we see the same pattern of behaviour seen for the evolution of $%
\alpha $ in the BSBM theory$.$ The smaller the curvature term, so the longer
the dust-dominated era lasts, and the greater the fall in the value of $G$,
and the smaller its ultimate asymptotic value when the curvature intervenes
to turn off the variation. In general, in such cosmologies, if there exists
a critical value of $G$ below which living complexity cannot be sustained,
then a universe that is too close to flatness will have a smaller interval
of cosmic history during which it can support life.

So far, we have discussed only the independent variation of $\alpha $ and $G$%
. What happens if they both vary at the same time? Previous studies of
varying constants have only examined the time-variation of a single
'constant'. We have produced a unified theory \cite{bms}, which incorporates
the BSBM varying $\alpha $ and BD varying $G$ theories discussed above. When
both $\alpha $ and $G$ are allowed to vary simultaneously in this theory we
find \cite{bms} that our general conclusions still hold, although the
quantitative details are changed. During the dust era of a flat Friedmann
universe with varying $\alpha (t)$ and $G(t),$their time-evolution
approaches an attractor in which the product $\alpha G$ is a constant and

\begin{equation}
\alpha \propto G^{-1}\propto t^n  \label{2}
\end{equation}
where $n$ is given by eq. (\ref{n}). Thus we see that the $G$ evolution is
left unchanged by the effects of varying $\alpha $, but variation of $G$
changes the time evolution of $\alpha (t)$ from a logarithm to a power-law
in time. As before, the longer the dust era lasts before it is ended by
deviation from flatness or zero cosmological constant, the longer the
time-increase of $\alpha $ continues, inevitably leading to values that make
any atom-based complexity impossible.

\section{ Discussion}

We have shown that some theories which include the time variation
of traditional constants like $\alpha $ and $G$ introduce
significant new anthropic considerations. A theory which
self-consistently introduces the space-time variation of a
traditional constant scalar quantity is strongly constrained in
form by the requirements of causality and second-order propagation
equations \cite{bek2}. Typically, this requirement leads to
equations for the driving scalar, $\varphi $ that have the form
$\Box \varphi $ proportional to linear combinations of the
energy-momentum components. Explicit examples are provided by the
Bekenstein-Sandvik-Barrow-Magueijo and Brans-Dicke theories. This
structure ensures that the evolution of the 'constant' whose
variations are derived from those of $\varphi $ is strongly
dependent upon the material or geometrical source governing the
background expansion dynamics. In the case of varying $\alpha $ we
have shown elsewhere \cite{bsm}, \cite{sbm} that this ties the
epoch after which time-variations in $\alpha $ become very small
to the time when the cosmological constant starts to accelerate
the expansion of the universe. In these theories there is
therefore the possibility of a habitable time zone of finite
duration during which a constant like $\alpha $ or $G$ falls
within a biologically acceptable range.

Surprisingly, there has been almost no consideration of habitability in
cosmologies with time-varying constants since Haldane's discussions \cite
{hal} of the biological consequences of Milne's bimetric theory of gravity
with two timescales, one for atomic phenomena, another for gravitational
phenomena \cite{mil}. Since then attention has focussed upon the
consequences of universes in which the constants are different but still
constants. Those cosmologies with varying constants that have been studied
have not considered the effects of curvature or $\Lambda $ domination on the
variation of constants and have generally considered power-law variation to
hold for all times. The examples described here show that this restriction
has prevented a full appreciation of the coupling between the expansion
dynamics of the universe and the values of the constants that define the
course of local physical processes within it. Our discussion of a theory
with varying $\alpha $ shows for the first time a possible reason why the
3-curvature of universes and the value of any cosmological constant may need
to be bounded {\it below} in order that the universe permit atomic life to
exist for a significant period. Previous anthropic arguments have shown that
the spatial curvature of the universe and the value of the cosmological
constant must be bounded {\it above} in order for life-supporting
environments (stars) to develop. We note that the lower bounds discussed
here are more fundamental than these upper bounds because they derive from
changes in $\alpha $ which have direct consequences for biochemistry whereas
the upper bounds just constrain the formation of astrophysical environments
by gravitational instability (for alternative scenarios see ref. \cite{ag}).
Taken together, these arguments suggest that within an ensemble of all
possible worlds where $\alpha $ and $G$ are time variables, there might only
be a finite interval of {\it non-zero }values of the curvature and
cosmological constant contributions to the dynamics that both allow galaxies
and stars to form and their biochemical products to persist.

\paragraph{Acknowledgements}

We would like to thank Bernard Carr, Andrei Linde, and Martin Rees
for discussions.  HBS acknowledges support from the Research
Council of Norway. The numerical work shown herein was performed
on COSMOS, the Origin 2000 supercomputer owned by the UK-CCC.

\end{document}